\documentclass[letterpaper, 10 pt, conference]{ieeeconf}  % Comment this line out
\IEEEoverridecommandlockouts                              % This command is only
\usepackage{amsmath}
\usepackage{mathtools}
\usepackage{amssymb}
\usepackage{gensymb}
\usepackage{bbold}
\usepackage[dvipsnames, usenames]{xcolor}
\usepackage{enumerate}
\usepackage{comment}
\usepackage{soul}
\usepackage{subfigure}
\usepackage{pdfpages}
\usepackage[hidelinks]{hyperref}

\newcommand{\RN}{\mathbb{R}} 				% % Set of Real Numbers

\newcommand{\BR}[1]{\left({ #1 }\right)}	% % Define Brace
\newcommand{\CBR}[1]{\left\lbrace #1 \right\rbrace}
\newcommand{\norm}[1]{\left\Vert #1\right\Vert}	% % Define Norm
\newcommand{\ABS}[1]{\lvert #1 \rvert}
\newcommand{\PARDIFF}[2]{\ensuremath{\frac{\partial #1}{\partial #2}}}

\newcommand{\DEL}[1]{\unskip}
\newcommand{\SPAN}[1]{\textbf{span}\CBR{ #1 }}
\newcommand{\NULL}[1]{\textbf{Null}\BR{ #1 }}
\newcommand{\COL}[1]{\textbf{Col}\BR{ #1 }}
\newcommand{\RANK}[1]{\textbf{rank}\BR{ #1 }}

\newcommand{\DIAG}[1]{\textbf{diag}\BR{ #1 }}
\newcommand{\BLKDIAG}[1]{\textbf{blkdiag}\BR{ #1 }}
\newcommand{\CARD}[1]{\textbf{card}\BR{ #1 }}
\newtheorem{remark}{Remark}
\newtheorem{theorem}{Theorem}
\newtheorem{lemma}{Lemma}
\newtheorem{proposition}{Proposition}

\title{\LARGE \bf
Angle-Constrained Formation Control for Circular Mobile Robots
}
\author{Nelson P.K. Chan$^{\star}$, Bayu Jayawardhana$^{\star}$, and Hector Garcia de Marina$^{\dagger}$
\thanks{
    {The work of N.P.K. Chan \& B. Jayawardhana is supported by the Region of Smart Factories (ROSF) project financed by REP-SNN and by the STW Smart Industry 2016 programme. 
    The work of H.G. de Marina is supported by the grant \emph{Atraccion de Talento} 2019-T2/TIC-13503 from the Government of the Autonomous Community of Madrid.}
    }
\thanks{$^{\star}$N.P.K. Chan and B. Jayawardhana are with 
    Engineering and Technology Institute Groningen, 
    University of Groningen, 
    Nijenborgh 4, 9747AG,
    Groningen, the Netherlands
    (email: {\tt\small n.p.k.chan@rug.nl, b.jayawardhana@rug.nl})
    }%
\thanks{$^{\dagger}$H.G. de Marina is with 
    the Department of Computer Architecture and Automatic Control 
    at the Faculty of Physics, 
    Universidad Complutense de Madrid{, 28040, Madrid, Spain}
    (email: {\tt\small hgarciad@ucm.es})
    }%
}

\begin{document}

% % Make Title
\maketitle
\thispagestyle{empty}
\pagestyle{empty}

%%%%%%%%%%%%%%%%%%%%%%%%%%%%%%%%%%%%%%%%%%%%%%%%%%%%%%%%%%%%%%%%%%%%%%%%%%%%%%%%
\begin{abstract}
In this letter, we investigate the formation control problem of mobile robots moving in the plane where, instead of assuming robots to be simple points, each robot is assumed to have the form of a disk with equal radius. 
Based on interior angle measurements of the neighboring robots' disk, which can be obtained from low-cost vision sensors, we propose a gradient-based distributed control law and show the exponential convergence property of the associated error system. 
By construction, the proposed control law has the appealing property of ensuring collision avoidance {between neighboring robots}.
We also present simulation results for {a team} of four circular mobile robots forming a rectangular shape.
\end{abstract}

\section{INTRODUCTION} \label{sec:Introduction}
Formation control studies the problem of controlling the spatial deployment of teams of mobile robots in order to achieve a specific geometric shape.
By maintaining a certain geometric shape, the teams can subsequently be deployed to perform complex missions.
Recent advances in this field focus on the design of distributed algorithms such that the formation control problem can be solved by only exploiting local information. 

Over the years, different approaches for formation control {have} been studied, and these can be classified according to the sensing and control variables that can be related to a geometrical property of the desired deployment for the robots \cite{Oh2015}. 
{One} class of formation control strategies {is}  the rigidity-based control strategies. 
{In this class,} rigidity theory plays a key role in characterizing a (at least locally) {unique target deployment, which can be achieved by a systematic design of distributed control laws}. 
Utilizing the \textit{distance} \cite{Oh2015, Sun2016} \textit{(or bearing} \cite{Zhao2016, Zhao2019}) \textit{rigidity theory}, we can define a specific deployment or \textit{target formation shape} in terms of a set of inter-robot distance (or bearing) constraints. 
The robots use available relative position or distance (or relative bearing) measurements in the design and execution of the distributed control law{s}. 
Recently, new rigidity theories, such as \textit{angle rigidity} \cite{Chen2019}, \textit{ratio-of-distance-rigidity} \cite{Cao2018}, and \textit{bearing-ratio-of-distance-rigidity theory} \cite{Cao2019} {have also been}  developed for characterizing a (at least locally) unique target formation shape using {a set of}  angle, ~{ratio-of-distance}, and ~{bearing-ratio-of-distance constraints}{, respectively}. 
{These theories focus on providing more flexibility to the target deployment by allowing scaling or rotational motions.}

{In addition, several works deal with practical aspects when implementing the proposed rigidity-based control strategies in real world settings. 
Among others, \cite{Mehdifar2019} considers robust distance-based formation control with prescribed performance, taking into account collision avoidance and connectivity maintenance between neighboring agents while they are subjected to unknown external disturbances; \cite{Frank2018} considers the bearing-only formation control problem with limited visual sensing while \cite{Marina2015} introduces estimators for controlling distance rigid formations under inconsistent measurements. 
}

One common aspect in the above-mentioned rigidity-based formation control theories is that the mobile robots are assumed to be simple points. 
As each robot is represented by a point in the plane, there can be only one relative position, distance, or bearing {measurement} between a pair of neighboring mobile robots. 
Instead of treating each robot as a point, we treat robots in this work as objects with area so that multiple features in the area can be measured by its neighbors. 
In particular, we assume each mobile robot to have a circular shape and move with single-integrator dynamics in the plane.
Furthermore, each mobile robot can observe \textit{two} distinctive features from its designated neighboring robots.
These are the outermost points of the neighboring robots' {disk} that can be seen from its centroid. 
In other words, we have the internal angle information of the neighboring robots.  
The desired formation shape can then be described in terms of feasible internal angle constraints, which  have a close relationship to the distance constraints {that are} used in distance-based formation control. 
This approach enables us to make the following novel contribution in the field of formation control:
\textit{
{
We provide an angle-constrained formation control algorithm, which resembles distance-based formation control. 
The main feature of our algorithm is that it requires only direction/bearing/unit vectors as measurements instead of a vector (that requires range and direction).
Furthermore, our algorithm provides collision avoidance guarantees where the clearance distance (which is twice the radius) between neighboring robots is not breached by design.
}
}

This letter is organized as follows.
Section \ref{sec:Preliminaries} provides  preliminaries on graph and distance rigidity theory. 
Next, in Section \ref{sec:ProblemSetup}{,} the problem setting and  problem formulation are presented. 
Section \ref{sec:grad} provides details concerning the control design and the local exponential convergence of the {associated} error dynamics. 
A numerical example is included in Section \ref{sec:NumericalExample} and 
{Section \ref{sec:Conclusions} concludes our work.}

\textit{Notation.}
The cardinality of a given set $ \mathcal{S} $ is denoted by $ \CARD{\mathcal{S}} $.
For a vector $ x \in \RN^{n} $, 
$ x^{\top} $ is the transpose, $ x^{\perp} $ is the perpendicular vector 
satisfying $ x^{\top} x^{\perp} = 0 = {\BR{x^{\perp}}}^{\top} x $, and $ \norm{x} = \sqrt{x^{\top}x} $ is the $ 2 $-norm of $ x $. 
The vector $ \mathbb{1}_{n} $ (or $ \mathbb{0}_{n} $) denotes the vector with entries being all $ 1 $s (or $ 0 $s).
The set of all combinations of linearly independent vectors $ v_{1}, \, \dots, \, v_{k} $ is denoted by $ \SPAN{v_{1}, \, \dots, \, v_{k}} $. 
For a matrix $ A \in \RN^{m \times n} $, {$ \NULL{A} \subset \RN^{n} $, $ \COL{A} \subset \RN^{m} $,} and $ \RANK{A} $ denotes its null space, its column space, and its rank{, respectively}.
The $ n \times n $ identity matrix is denoted by $ I_{n} $ while $ \DIAG{v} $ (or $ \BLKDIAG{A_{1}, \, \dots , \, A_{k}} $) is the diagonal (or block diagonal) matrix with entries of vector $ v $ (or matrices $ A_{1}, \, \dots, \, A_{k} $) on the main diagonal (or block).
Finally, given matrices $ A \in \RN^{m \times n} $ and $ B \in \RN^{p \times q} $, $ A \otimes B \in \RN^{mp \times nq} $ is the Kronecker product of $ A $ and $ B ${,} and we denote $ \widetilde{A} = A \otimes I_{d} \in \RN^{md \times nd} $. 

\section{PRELIMINARIES} \label{sec:Preliminaries}
{
This section provides the necessary concepts in graph theory and distance rigidity theory.
For a more detailed exposure of the material, we refer to for instance, \cite{Bullo2019} on graph theory, and \cite{Anderson2008, Queiroz2019} on distance rigidity theory.
}
\subsection{Graph theory}
An \textit{undirected graph} $ \mathcal{G} $ is defined as a pair $ \BR{\mathcal{V}, \, \mathcal{E}} $, where $ \mathcal{V} \coloneqq \CBR{ 1, \: 2, \: \dots, \: n} $ and $ \mathcal{E} \coloneqq \CBR{\CBR{i, \, j} \, \rvert \, i, \, j \in \mathcal{V}} $ denote the finite set of \textit{vertices} and the set of unordered pairs $ \CBR{i, \, j} $ of the vertices, called \textit{edges}. 
We assume the graph does not have self-loops, i.e., $ \CBR{i, \, i} \not\in \mathcal{E}, \, \forall i \in \mathcal{V} $, and $ \CARD{\mathcal{E}} = m $.
The edge $ \CBR{i, \, j} $ indicates vertices $ i $ and $ j $ are \textit{neighbors of each other}. 
The set of neighbors of vertex $ i $ is denoted by $ \mathcal{N}_{i} \coloneqq \CBR{ j \in \mathcal{V} \, \rvert \, \CBR{i, \, j} \in \mathcal{E}} $. 
By assigning an arbitrary orientation to each edge of $ \mathcal{G} $, we obtain an \textit{oriented} graph $ \mathcal{G}_{\text{orient}}$. 
The incidence matrix $ H \in \CBR{0, \, \pm 1}^{m \times n} $ associated to $ \mathcal{G}_{\text{orient}} $ has rows encoding the $ m $ oriented edges and columns encoding the $ n $ vertices. 
$ \left[H\right]_{ki} = \BR{{-1}} +1 $ if vertex $ i $ is the (tail) head of edge $ k $ and $ \left[H\right]_{ki} = 0 $ otherwise.
For a connected and undirected graph, we have $ \NULL{H} = \SPAN{\mathbb{1}_{n}}$ and $ \RANK{H} = n - 1 $. 

\subsection{Distance rigidity theory}
Let 
$ p_{i} = \left[x_{i}, \, y_{i} \right]^{\top} \in \RN^{2} $ 
be a point in the plane and a collection of points, called a \textit{configuration}, be given by the stacked vector $ p = {\begin{bmatrix} p_{1}^{\top} & \cdots & p_{n}^{\top} \end{bmatrix}}^{\top} \in \RN^{2n} $.
We can embed the graph $ \mathcal{G}\BR{\mathcal{V}, \: \mathcal{E}} $ into the plane by assigning to each vertex $ i \in \mathcal{V} $, a point $ p_{i} \in \RN^{2} $.
The pair $ \mathcal{F}_{p} \coloneqq \BR{\mathcal{G}, \: p} $ denotes a \textit{framework} in $ \RN^{2} $.
We assume $ p_{i} \neq p_{j} $ if $ i \neq j $, i.e., no two vertices are mapped to the same position. 
Related to $ \mathcal{F}_{p} $, we define the \textit{distance rigidity function} $ r_{\text{dist}} : \RN^{2n} \to \RN_{>0}^{m} $ as 
\begin{equation} \label{eq:RigidityFunction}
    r_{\text{dist}}\BR{p} \coloneqq \frac{1}{2}
        {\begin{bmatrix}
            \cdots
            & 
            {\norm{p_{j} - p_{i}}}^{2}
            & 
            \cdots 
        \end{bmatrix}}^{\top}
        , \: \forall \, \CBR{i, \, j} \in \mathcal{E},
\end{equation}
with each entry of the vector being half the squared distance between two points. 
Given the distance rigidity function \eqref{eq:RigidityFunction}, we say a framework $ \mathcal{F}_{p} $ is \textit{distance rigid}, if there exists a neighborhood $ \mathcal{U}_{p} $ of $ p $ such that, if $ q \in \mathcal{U}_{p} $ and $ r_{\text{dist}}\BR{p} = r_{\text{dist}}\BR{q} $, then $ \mathcal{F}_{q} $ is congruent to $ \mathcal{F}_{p} $.
Let $ z_{ij} = p_{j} - p_{i} \in \RN^{2} $ be the relative position vector associated to $ \CBR{i, \, j} \in \mathcal{E} $, and  $ z \in \RN^{2m} $ be the stacked vector of $ z_{ij} $s. 
Using the incidence matrix $ \widetilde{H} \in \RN^{2m \times 2n} $, we obtain $ z = \widetilde{H}p $.
{Besides,} let $ Z\BR{z} = \BLKDIAG{\CBR{z_{ij}}_{\CBR{i, \, j} \in \mathcal{E}}} \in \RN^{2m \times m} $. 
Using these expressions, \eqref{eq:RigidityFunction} can be written in compact form as $ r_{\text{dist}}\BR{p} = \frac{1}{2} Z^{\top}\BR{z} z $.
By taking the Jacobian of \eqref{eq:RigidityFunction}, we obtain the \textit{distance rigidity matrix} $ R_{\text{dist}}\BR{p} $ as 
\begin{equation} \label{eq:RigidityMatrix}
    R_{\text{dist}}\BR{p} \coloneqq \PARDIFF{r_{\text{dist}}\BR{p}}{p}  = Z^{\color{blue}{\top}}\BR{z} \widetilde{H} \in \RN^{m \times 2n}.
\end{equation}
{Let $ \delta p \in \RN^{2n} $ be an {infinitesimal} variation of $ p $. A motion $\delta p$ is said to be \textit{trivial} if $R_{\text{dist}}\BR{p} \delta p = \mathbb{0}_{m} $  corresponds to a translation and or a rotation of the entire framework}. 
Trivial motions in the plane are {a} translation in the $ x $- and in the $ y $-direction, a rotation, {and the combination thereof}, all applied to the entire framework.
We say a framework $ \mathcal{F}_{p} $ is \textit{infinitesimally distance rigid} if and only if the set of infinitesimally distance motions consists of only the trivial motions. 
This can be translated to the following condition on the distance rigidity matrix: $ \RANK{R_{\text{dist}}\BR{p}} = 2n - 3 $.
{Furthermore}, an infinitesimally distance rigid framework must have at least $ 2n - 3 $ edges.
If the number of edges $ m $ is exactly $ 2n - 3 $, then the framework is said to be \textit{minimally and infinitesimally distance rigid}. 

\begin{figure}
    \centering
    \includegraphics[width=0.495\textwidth]{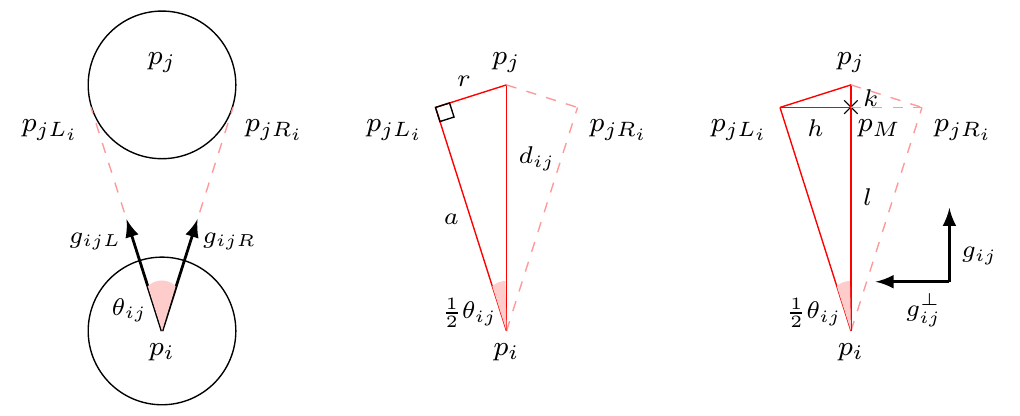}
    \caption{Sensing setup with robot $ i $ being the `observer' and robot $ j $ the `observed' robot. On the left panel, robot $ i $ detects the points $ p_{jL_{i}} $ and $ p_{jR_{i}} $ of robot $ j $ and the internal angle $ \theta_{ij} $ can be obtained from the bearing measurements $ g_{ijL} $ and $ g_{ijR} $. In the middle panel, we use geometrical arguments to relate $ \theta_{ij} $ to the inter-center distance $ d_{ij} $ and the radius $ r $. On the right panel, we have a geometrical view supporting Proposition \ref{prop:RelBearGradLaw}.}
    \label{fig:Sensing-Setup}
\end{figure}

\section{PROBLEM SETUP} \label{sec:ProblemSetup}
We consider a group of $ n $ mobile robots moving in the plane. 
Let $ \mathcal{V} = \CBR{1, \, 2, \, \dots, \, n} $ be the index set of the robots. 
Each robot has a circular shape with center specified by $ p_{i} \in \RN^{2} $ and radius by $ r_{i} \in \RN_{>0} $. 
For simplicity, we assume the {radii} of the robots have the same value and let $ r \in \RN_{>0} $ represent this common value. 
We assume the robots are moving with single-integrator dynamics, i.e.,
\begin{equation} \label{eq:AgentDynamics}
    \dot{p}_{i} \BR{t} = u_{i} \BR{t}, \: \forall \, i \in \mathcal{V},
\end{equation}
where $ u_{i} \in \RN^{2} $ is the controlled velocity to be designed. 
The group dynamics is given by $ \dot{p}\BR{t} = u\BR{t} $ with the stacked vectors $ p = {\begin{bmatrix} p_{1}^{\top} & \cdots & p_{n}^{\top} \end{bmatrix}}^{\top} \in \RN^{2n} $ and $ u = {\begin{bmatrix} u_{1}^{\top} & \cdots & u_{n}^{\top} \end{bmatrix}}^{\top} \in \RN^{2n} $. 

Each robot is equipped with a sensory system mounted at the center $ p_{i} $ of the robot. 
With the equipped sensory system, we assume the robots are able to detect two points on the surface of each of its designated neighbors.
To illustrate this, let us consider without loss of generality a pair of robots labeled $ i $ and $ j $ within the group of robots, see Fig. \ref{fig:Sensing-Setup}. 
We assume robot $ i $ has the role of `observer' and robot $ j $ is the `observed' robot. 
Since robot $ i $ is the observer, it is able to detect two points on the surface of robot $ j $. 
We denote the position of the detected points as $ p_{jL_{i}} $ and  $ p_{jR_{i}} $ to indicate these are the positions of robot $ j $ as detected by robot $ i $. 
The measurements from robot $ j $ that {are} available to robot $ i $ are the \textit{relative bearing measurements}
$ g_{ijL} = \frac{z_{ijL}}{\norm{z_{ijL}}} $ and $ g_{ijR} = \frac{z_{ijR}}{\norm{z_{ijR}}} $, with $ z_{ijL} = p_{jL_{i}} - p_{i} $ and $ z_{ijR} = p_{jR_{i}} - p_{i} $ being the relative position from the detected points to the center of robot $ i $. 
The two bearing vectors form an angle $ \theta_{ij} $ centered at $ p_{i} $, as can be seen in Fig. \ref{fig:Sensing-Setup}. 
By the inner product rule, we obtain
\begin{equation} \label{eq:CosineMeasurements}
    \cos \theta_{ij} = g_{ijL}^{\top} g_{ijR}
    .
\end{equation}

\begin{remark} \label{rem:GeometricView}
    It should be noted the lines in the direction of the unit vectors $ g_{ijL} $ and $ g_{ijR} $ are both tangent lines from the point $ p_{i} $ to robot $ j $. 
    Hence, these lines are perpendicular to the radius of the circle, i.e,
    $ \BR{p_{jL_{i}} - p_{j}} \perp z_{ijL} $ and $ \BR{p_{jR_{i}} - p_{j}} \perp z_{ijR} $.
    {Furthermore}, the triangle $ \Delta p_{j} p_{jL_{i}} p_{i} $ with vertices $ p_{j} $, $ p_{i} $, and $ p_{jL_{i}} $ and the triangle $ \Delta p_{j} p_{jR_{i}} p_{i} $ with vertices $ p_{j} $, $ p_{i} $, and $ p_{jR_{i}} $ are reflections of each other with the line connecting $ p_{j} $ and $ p_{i} $ as the line of reflection. 
    Hence, the angle $ \angle p_{jL_{i}} p_{i} p_{j} = \angle p_{j} p_{i} p_{jR_{i}} = \frac{1}{2} \theta_{ij} $. 
\end{remark}

\vspace{0.5\baselineskip}
By considering the geometry, we can obtain an alternative expression for $ \cos \theta_{ij} $, which is related to the {radii} of  and the inter-center distance between the robots. 
To this end, we first define some auxiliary relative state variables. 
For robots $ i $ and $ j $,
let $ z_{ij} = p_{j} - p_{i} $ denotes the relative position, $ d_{ij} = \norm{z_{ij}} $ the distance, and $ g_{ij} = \frac{z_{ij}}{d_{ij}} $ the relative bearing between the centers of the robots.
{Also,} $ g_{ij}^{\perp} $ is the perpendicular vector obtained by rotating $ g_{ij} $ counterclockwise by $ 90\degree $. 
We have $ g_{ij}^{\perp} = J g_{ij} $ with $ J \coloneqq \left[\begin{smallmatrix} 0 & -1 \\ 1 & 0 \end{smallmatrix} \right]$ being the rotation matrix. 

\vspace{0.5\baselineskip}
\begin{proposition} \label{prop:CosineGeometry}
    The internal angle $ \theta_{ij} $ is related to the inter-center distance $ d_{ij} $ between the robots $ i $ and $ j $ and the {radii} $ r $ of the robots as 
    \begin{equation} \label{eq:CosineRobot_ij}
        \cos \theta_{ij} = 1 - 2{\BR{\frac{r}{d_{ij}}}}^{2}.
    \end{equation}
\end{proposition}

\vspace{0.5\baselineskip}
\begin{proof}
    The desired result can be obtained by employing the cosine double-angle identity $ \cos \alpha = 1 - 2 \sin^{2} \frac{1}{2}\alpha $ and noting from Remark \ref{rem:GeometricView} that $ \Delta p_{j} p_{jL_{i}} p_{i} $ is a right triangle with $ \sin \frac{1}{2} \theta_{ij} = \frac{r}{d_{ij}} $.
    Fig. \ref{fig:Sensing-Setup} provides the geometric illustration.
\end{proof}

\vspace{0.5\baselineskip}
Note that \eqref{eq:CosineMeasurements} and \eqref{eq:CosineRobot_ij} are equivalent for obtaining the internal angle $ \theta_{ij} $; the former is based on the available bearing measurements while the latter is based on geometry. 

\vspace{0.5\baselineskip}
\begin{remark} \label{rem:FeasibleRegionCosine}
    As robot{s} $ i $ and $ j $ are of circular shape, the feasible interval for the inter-center distance $ d_{ij} $ is $ d_{ij}^{\text{feas}} \in \BR{2r, \infty} $. 
    This {also} poses  restrictions on the value for $ \theta_{ij} $ and $ \cos \theta_{ij} $. 
    From \eqref{eq:CosineRobot_ij}, it follows that $ d_{ij}^{\text{feas}} \in \BR{2r, \infty} $ implies $ \cos \theta_{ij}^{\text{feas}} \in \BR{\frac{1}{2}, \, 1} $ and $ \theta_{ij}^{\text{feas}} \in \BR{0, \, 60\degree} $.
    Correspondingly, an increase in the value of $ d_{ij} $ results in an increase of $ \cos \theta_{ij} $ and a decrease of $ \theta_{ij} $.
\end{remark}

We can rewrite \eqref{eq:CosineRobot_ij} as $ d_{ij} = \sqrt{\frac{2r^{2}}{1 - \cos \theta_{ij}}}$. 
By obtaining $ \cos \theta_{ij} $ from \eqref{eq:CosineMeasurements} and knowing $ r $, we  can infer the inter-center distance $ d_{ij} $. 
With this observation, we  define an \textit{internal angle rigidity function} $ r_{\text{angle}} \: : \RN^{2n} \to \RN_{>0}^{m} $ given by 
\begin{equation} \label{eq:AngleRigidityFunction}
    r_{\text{angle}}\BR{p} = {\begin{bmatrix} \cdots  & \cos \BR{\theta_{ij}} & \cdots \end{bmatrix}}^{\top}, \, \forall \, \CBR{i, \, j} \in \mathcal{E}
\end{equation}
for describing a framework $ \mathcal{F}_{p}\BR{\mathcal{G}, \, p} $.
By Remark \ref{rem:FeasibleRegionCosine}, there is a one-to-one relationship between the newly defined rigidity function \eqref{eq:AngleRigidityFunction} and the distance rigidity function \eqref{eq:RigidityFunction}.
The Jacobian of \eqref{eq:AngleRigidityFunction} is
\begin{equation} \label{eq:Jacob-ARF}
    \begin{aligned}
        R_{\text{angle}}\BR{p} 
            & 
            = \PARDIFF{r_{\text{angle}}\BR{p}}{p}  
            = \PARDIFF{r_{\text{angle}}\BR{p}}{{q}}  \PARDIFF{{q}}{p}
            = D\BR{d} R_{\text{dist}}
            ,
    \end{aligned}
\end{equation}
with $ d \in \RN^{m} $ being the {stacked} vector of  distances $ d_{ij} $s, {$ q = \BR{\DIAG{d}d} \in \RN^{m} $,} and 
{$ D\BR{d} = 4r^{2} \DIAG{\CBR{d_{ij}^{-4}}_{\CBR{i, \, j} \in \mathcal{E}}} \in \RN^{m \times m} $}. 
The matrix $ D\BR{d} $ is positive definite as each $ d_{ij} > 2r > 0 $. Thus{,} we have $ \RANK{R_{\text{angle}}} = \RANK{R_{\text{dist}}} $.

Now we can define the desired target formation shape by a framework $ \mathcal{F}_{p^{\star}}\BR{\mathcal{G}, \, p^{\star}} $ where the vector $ p^{\star} \in \RN^{2n} $ satisfies a set of desired internal angle constraints $ r_{\text{angle}}\BR{p^{\star}} $.  
One way to obtain the internal angle constraints is to employ \eqref{eq:CosineRobot_ij} when the desired distance constraints {are given}.
{Moreover,} the formation $ \mathcal{F}_{p^{\star}} $ is \textit{minimally} and \textit{infinitesimally} rigid in the distance rigidity sense.
The formation control problem that is considered in this work can be formulated as follows:

\vspace{0.5\baselineskip}
\noindent
\textbf{Angle-constrained Formation Control Problem with Collision Avoidance: } 
Given a set of feasible internal angle constraints\footnote{We give a formal definition of such a set in Section \ref{sec: shapes}.} $ \CBR{\theta_{ij}^{\star}}_{\CBR{i, \, j} \in \mathcal{E}} $ obtained using \eqref{eq:CosineRobot_ij} from a minimally and infinitesimally rigid framework $ \mathcal{F}_{p^{\star}} $ and {an}  initial {configuration} $ p\BR{0} \in \RN^{2n} $ with $ \norm{p_{j}\BR{0} - p_{i}\BR{0}} > 2r, \forall \, \CBR{i, \, j} \in \mathcal{E} $.
Design a control law $ u_{i}\BR{t}, \, \forall \, i \in \mathcal{V} $ utilizing only the neighboring measurements obtained as in \eqref{eq:CosineMeasurements} such that $ \forall \CBR{i, \, j} \in \mathcal{E} $
\begin{itemize}
    \item {\it Collision avoidance:} $ \norm{p_{j}\BR{t} - p_{i}\BR{t}} > 2r $, $ \forall t \geq 0 $;
    
    \item {\it Convergence:} $ \theta_{ij}\BR{t} \rightarrow \theta_{ij}^{\star} $ as $ t \rightarrow \infty $.
\end{itemize}

\section{GRADIENT-BASED CONTROL DESIGN} \label{sec:grad}
In this section, we pursue a gradient-based control design approach utilizing angle-based potential functions for solving the formation control problem. 
To each edge $ \CBR{i, \, j} \in \mathcal{E} $, we  define the error signal $ e_{ij}\BR{t} = \cos \theta_{ij}\BR{t} - \cos \theta_{ij}^{\star} $.
By Remark \ref{rem:FeasibleRegionCosine}, we deduce the feasible region for the error signal {is} $ e_{ij}^{\text{feas}} \in \BR{-c_{ij}, \, f_{ij}} ${,} with $ c_{ij} = \cos \theta_{ij}^{\star} - \frac{1}{2} $ and $ f_{ij} = 1 - \cos \theta_{ij}^{\star} $.
Both $ c_{ij} $ and $ f_{ij} $ are strictly positive. 

\subsection{Proposed angle-based potential function}
{For a robot pair $ \CBR{i, \, j} $, we} take as potential function 
\begin{equation} \label{eq:LocalPotentialFunction}
    V_{ij}\BR{e_{ij}} = \frac{1}{2} r{\BR{\frac{\cos \theta_{ij} - \cos \theta_{ij}^{\star}}{\cos \theta_{ij} - \frac{1}{2}}}}^{2} = \frac{1}{2} r {\BR{\frac{e_{ij}}{e_{ij} + c_{ij}}}}^{2}.
\end{equation}
The denominator term $ \cos \theta_{ij} - \frac{1}{2} $ ensures collision avoidance between the {neighboring} robots $ i $ and $ j $, i.e., $ \norm{p_{j}\BR{t} - p_{i}\BR{t}} > 2r, \, \forall t>0 $ given that $ \norm{p_{j}\BR{0} - p_{i}\BR{0}} > 2r $.
The function $ V_{ij}\BR{e_{ij}} $ is non-negative in $ e_{ij}^{\text{feas}} $.
Furthermore, $ V_{ij}\BR{e_{ij}} = 0 $ if and only if $ e_{ij} = 0 $ and $ V_{ij}\BR{e_{ij}} \rightarrow \infty $ if {$ e_{ij} $ approaches the lower bound $ -c_{ij} $ from above,} i.e., when the mobile robots are approaching each other.

The first derivative $ v_{ij}\BR{e_{ij}} \coloneqq \PARDIFF{}{e_{ij}} V_{ij}\BR{e_{ij}} $ can be obtained as 
$ v_{ij}\BR{e_{ij}} = r \frac{e_{ij} c_{ij}}{{\BR{e_{ij} + c_{ij}}}^{3}} $.
The value of $ v_{ij}\BR{e_{ij}} $ equals zero if and only if $ e_{ij} = 0 $ and the sign of $ v_{ij} $ depends on the sign of $ e_{ij} $.

The second derivative $ k_{ij}\BR{e_{ij}} \coloneqq \PARDIFF{^{2}}{e_{ij}^{2}} V_{ij}\BR{e_{ij}} $ is given as 
~$ k_{ij}\BR{e_{ij}} = r\frac{c_{ij}}{\BR{e_{ij} + c_{ij}}^{4}} \BR{-2e_{ij} + c_{ij}} $.
 $ k_{ij}\BR{e_{ij}} $ is positive when $ e_{ij} < \frac{1}{2} c_{ij} $. 
Recall $ e_{ij}^{\text{feas}} \in \BR{-c_{ij}, \, f_{ij}} $; {therefore, we need} to determine when $ \frac{1}{2} c_{ij} \lesseqqgtr f_{ij} $. 
By some algebraic computations, we obtain $ \frac{1}{2} c_{ij} \lesseqqgtr f_{ij} $ if and only if 
$ \cos \theta_{ij}^{\star} \lesseqqgtr \frac{5}{6} $.
When $ \cos \theta_{ij}^{\star} < \frac{5}{6} $, we have the region for which $ k_{ij}\BR{e_{ij}} $ is positive is a subset of $ e_{ij}^{\text{feas}} $, whereas when $\cos \theta_{ij}^{\star} \geq \frac{5}{6} $, we have $ k_{ij}\BR{e_{ij}} $ is positive over the entire domain $ e_{ij}^{\text{feas}} $.

The properties of 
\eqref{eq:LocalPotentialFunction} {will be used later} for deriving the exponential convergence of the error dynamics.

\subsection{Gradient-based control law for each robot}
The local potential function for each robot $ i $ is  $ V_{i}\BR{e} = \sum_{j \in \mathcal{N}_{i}} V_{ij}\BR{e_{ij}} $ with 
$ e \in \RN^{m} $ being the stacked vector of error signals $ e_{ij} $s.
The control input $ u_{i}\BR{t} $ 
is then
\begin{equation} \label{eq:GradientControlLawDef}
    u_{i}\BR{t} 
        = - {\BR{\PARDIFF{}{p_{i}} V_{i}\BR{e}}}^{\top} 
        = - \sum_{j \in \mathcal{N}_{i}} {\BR{\PARDIFF{}{p_{i}} V_{ij}\BR{e_{ij}}}}^{\top}.
\end{equation}
Utilizing \eqref{eq:CosineRobot_ij}, the term $ \PARDIFF{}{p_{i}} V_{ij}\BR{e_{ij}} $ can be evaluated {as} 
\begin{equation} \label{eq:GradientControlLawExp}
    \begin{aligned}
        u_{ij}^{\top} \coloneqq 
        \PARDIFF{}{p_{i}} V_{ij} \BR{e_{ij}}
            = - v_{ij}\BR{e_{ij}} \frac{4r^{2}}{d_{ij}^{4}}z_{ij}^{\top}
            .
    \end{aligned}
\end{equation}

Note that \eqref{eq:GradientControlLawExp} requires relative state variables 
$ d_{ij} ${,} $ z_{ij} ${, and the knowledge of $ r $}.
However, robot $ i $ has access to {only} the relative bearing measurements $ g_{ijL} $ and $ g_{ijR} $ for each $ j \in \mathcal{N}_{i} $.
Nonetheless, we show that the gradient-control law \eqref{eq:GradientControlLawDef} can be implemented using these available measurements.

\vspace{0.5\baselineskip}
\begin{proposition} \label{prop:RelBearGradLaw}
  The gradient-based control law \eqref{eq:GradientControlLawDef} can be implemented by each robot $ i \in \mathcal{V} $ using the set of available measurements 
  $ \CBR{\CBR{g_{ijL}}_{j \in \mathcal{N}_{i}}, \, \CBR{g_{ijR}}_{j \in \mathcal{N}_{i}}} $.
\end{proposition}

\vspace{0.5\baselineskip}
\begin{proof}
    To implement \eqref{eq:GradientControlLawDef}, we need to rewrite \eqref{eq:GradientControlLawExp} in terms of the available measurements $ g_{ijL} $ and $ g_{ijR} $.
    To this end, first, we seek expressions for the positions $ p_{jL_{i}} $ and $ p_{jR{i}} $.
    Let us consider again Fig. \ref{fig:Sensing-Setup}.
    Denote the intersection between the lines connecting the center of the robots and the two intersection points as $ p_{M} $ {(marked with the $ \times $-symbol in the right panel of Fig. \ref{fig:Sensing-Setup})}. 
    Let $ \norm{p_{jL_{i}} - p_{M}} = h $
    , $ \norm{p_{j} - p_{M}} = k $, and $ \norm{p_{i} - p_{M}} = l $ satisfying $ k + l = d_{ij} $.
    $ l $ can {also} be written as a fraction of the inter-center distance $ d_{ij} $, i.e., $ l = s d_{ij} $ with $ s \in \BR{0, 1} $.
    We can now express the positions $ p_{jL_{i}} $ and $ p_{jR_{i}} $ as 
    $ p_{jL_{i}} = p_{j} - k g_{ij} + h g_{ij}^{\perp} $, and $ p_{jR_{i}} = p_{j} - k g_{ij} - h g_{ij}^{\perp} $.
    Recall $ g_{ij} $ is the unit vector between the centers of the robots. 
    Subsequently, the relative position $ z_{ijL} $ and $ z_{ijR} $ can be obtained as 
    $ z_{ijL} = l g_{ij} + h g_{ij}^{\perp} $, and $ z_{ijR} = l g_{ij} - h g_{ij}^{\perp} $,
    while their sum equals
    $
        z_{ij+} 
            = z_{ijL} + z_{ijR} 
            = 2 s z_{ij}.
    $
    Due to the reflection observation {in Remark \ref{rem:GeometricView}}, we have $ \norm{z_{ijL}} = \norm{z_{ijR}} = \sqrt{l^{2} + h^{2}} \eqqcolon a $.
    Using the previous computations, we obtain for the sum of the relative bearing measurements 
    $
            g_{ij+} 
                = g_{ijL} + g_{ijR} 
                = 2\frac{s}{a} z_{ij}.
    $
    In addition, 
    $
            \frac{g_{ij+}}{\norm{g_{ij+}}^{2}} 
                = \frac{2 \frac{s}{a} z_{ij}}{4 {\BR{\frac{s}{a}}}^{2} d_{ij}^{2}}  \iff 2 \frac{z_{ij}}{d_{ij}^{2}} 
                = 4 \frac{s}{a} \frac{g_{ij+}}{\norm{g_{ij+}}^{2}}
                .
    $
    {Since}  $ s = \frac{l}{d_{ij}} $, we can rewrite $ \frac{s}{a} $ as 
    $
            \frac{s}{a}
                = \frac{l}{d_{ij}a} \frac{r}{r}
                = \frac{1}{r} \sin \frac{1}{2} \theta_{ij} \cos \frac{1}{2} \theta_{ij}
                = \frac{1}{2r} \sin \theta_{ij} 
    $
    by using $ \sin \frac{1}{2} \theta_{ij} = \frac{r}{d_{ij}} $, $ \cos \frac{1}{2} \theta_{ij} = \frac{a}{d_{ij}} = \frac{l}{a} $, and the sine double-angle identity $ \sin 2\alpha = 2 \sin \alpha \cos \alpha $.
    Substituting the obtained expressions in \eqref{eq:GradientControlLawExp} and utilizing \eqref{eq:CosineRobot_ij} yield
    \begin{equation} \label{eq:GradLocalBear}
        \begin{aligned}
            u_{ij}^{\top}
                = - 2 \widehat{v}_{ij}\BR{e_{ij}} \BR{1 - \cos \theta_{ij}} \sin \theta_{ij}
                {\norm{g_{ij+}}^{-2} g_{ij+}}
                ,
        \end{aligned}
    \end{equation}
	where $ \widehat{v}_{ij}\BR{e_{ij}} = \frac{v_{ij}\BR{e_{ij}}}{r} {= \frac{e_{ij}c_{ij}}{{\BR{e_{ij} + c_{ij}}}^{3}}}$, {i.e, using \eqref{eq:GradLocalBear}, we can implement \eqref{eq:GradientControlLawDef} without knowledge of the range information and the radii of the robots}.
\end{proof}

\subsection{Gradient-based control law for the group of robots}

The overall potential function $ V\BR{e} $ can be expressed as the sum of all the individual potential functions $ V_{ij}\BR{e_{ij}} $, i.e., $ V\BR{e} = \sum_{\CBR{i, \, j} \in \mathcal{E}} V_{ij}\BR{e_{ij}} $.
The control law $ u_{i}\BR{t} $  in \eqref{eq:GradientControlLawDef} is then $ u_{i}\BR{t} = - {\BR{\PARDIFF{}{p_{i}} V\BR{e}}}^{\top} $.
By noting {$ \PARDIFF{}{p} V\BR{e} = \PARDIFF{V\BR{e}}{e} \PARDIFF{e}{{q}} \PARDIFF{{q}}{p} $,} 
we obtain the following compact form for the closed-loop formation {control} system:
\begin{equation} \label{eq:OverallClosedLoopExp}
    \dot{p}\BR{t} = - R_{\text{angle}}^{\top} v\BR{e},
\end{equation}
with the vector $ v\BR{e} \in \RN^{m} $ denoting the gradients of \eqref{eq:LocalPotentialFunction} for each {robot pair} $ \CBR{i, \, j} \in \mathcal{E} $. 

\vspace{0.5\baselineskip}
\begin{lemma} \label{lem:Prop-Form-Control}
    The closed-loop formation control system \eqref{eq:OverallClosedLoopExp} has the following properties:
    \begin{enumerate}[(1.)]
        \item The formation centroid $ p_{\text{cent}} = \frac{1}{n} \BR{\mathbb{1}_{n}^{\top} \otimes I_{2}} p $ is stationary, i.e., $ p_{\text{cent}}\BR{t} = p_{\text{cent}}\BR{0}, \, \forall t \geq 0 $;
        
        \item Each mobile robot can have its own local coordinate system for obtaining the required relative state measurements and implementing the desired control action.
    \end{enumerate}
\end{lemma}

\vspace{0.5\baselineskip}
\begin{proof}
    % The proof is straightforward and thus omitted.
    {The proof is similar to Lemma 4 in \cite{Sun2015}, and thus not provided here.}
\end{proof}

\subsection{Internal angle error system}
Using the definition of the error vector $ e $, and expressions \eqref{eq:CosineRobot_ij} and \eqref{eq:OverallClosedLoopExp}, we can obtain {the error}  dynamics 
\begin{equation} \label{eq:ErrorDynamics}
    \dot{e}\BR{t} 
        = \PARDIFF{e}{p} \dot{p} 
        = - R_{\text{angle}} R_{\text{angle}}^{\top} v\BR{e}
        = - F v\BR{e}
        .
\end{equation}
The matrix $ F = R_{\text{angle}} R_{\text{angle}}^{\top} = {{D\BR{d} R_{\text{dist}} R_{\text{dist}}^{\top} D\BR{d}}} \in \RN^{m \times m} $ is symmetric and at least positive semidefinite. 
{Moreover, for any infinitesimally and minimally distance rigid framework $ \mathcal{F}_{p^{\star}} $}, $ F $ can be shown to be a function of the error vector $ e $ {around the origin} by employing the law of cosines.
By this observation, we conclude the error dynamics given by \eqref{eq:ErrorDynamics} constitute an autonomous system. 

The main result will be the local exponential stability of the error dynamics \eqref{eq:ErrorDynamics}. 
To this end, we first construct a compact and invariant sub-level set for the overall potential function $ V\BR{e} $.
~\\
Previously, we have $ k_{ij}\BR{e_{ij}} > 0 $ holds if and only if $ e_{ij} < b_{ij} \coloneqq \min \CBR{\frac{1}{2} c_{ij}, \, f_{ij}}, \, \forall \, \CBR{i, \, j} \in \mathcal{E} $. 
Let $ b = \min \CBR{b_{ij}}_{\CBR{i, \, j} \in \mathcal{E}} > 0 $.
We define the `hypercube' as 
\begin{equation} \label{eq:HypercubeB}
    \mathcal{H}_{b} = \CBR{e \in \mathcal{CF} \, \rvert \, \ABS{e_{k}} < b, \, k \in \mathcal{K}}
    ,
\end{equation}
with $ \mathcal{CF} $ being the Cartesian product $ \BR{-c_{1}, \, f_{1}} \times \cdots \times \BR{-c_{m}, \, f_{m}} $ and $ \mathcal{K} = \CBR{1, \, \cdots, \, m} $ being the ordered edge index set.
Choose $ q \in \BR{0, b} $ such that
\begin{equation} \label{HyperballQ}
    \mathcal{B}_{q} = \CBR{e \in \mathcal{H}_{b} \, \rvert \, \norm{e} \leq q} \subset {{\mathcal{H}_{b}}}.
\end{equation}
Let $ \alpha = \min_{\norm{e} = q} V\BR{e} $.
As $ q \neq 0 $, we have $ V\BR{e} > 0 $ and also $ \alpha > 0 $.
Choose $ \beta \in \BR{0, \alpha} $ and define 
\begin{equation} \label{eq:OmegaBeta}
    \Omega_{\beta} = \CBR{e \in \mathcal{B}_{q} \, \rvert \, V\BR{e} \leq \beta}.
\end{equation}
By definition, the sub-level set $ \Omega_{\beta} $ is closed and as $ \Omega_{\beta} \subset \mathcal{B}_{q} $, it is also bounded. 
Thus, $ \Omega_{\beta} $ is a compact set. 
The time-derivative of $ V\BR{e} $ can be obtained as 
\begin{equation} \label{eq:OverallPotFuncDer}
    \begin{aligned}
        \dot{V}\BR{e} 
            = \PARDIFF{}{e} V\BR{e} \dot{e}
            = -v^{\top}\BR{e} F\BR{e} v\BR{e}
            \leq 0
            .
    \end{aligned}
\end{equation}
This implies $ V\BR{e\BR{t}} \leq V\BR{e\BR{0}} $.
Whenever $ e\BR{0} \in \Omega_{\beta} $, we have by \eqref{eq:OverallPotFuncDer} that $ e\BR{t} \in \Omega_{\beta} ${; therefore,}  the set $ \Omega_{\beta} $ is also positive invariant. 
As $ V\BR{e} \geq 0 $ and $ \dot{V}\BR{e} \leq 0 $, the overall potential function can serve as a candidate Lyapunov function.
We are ready to state and prove the main result. 

\vspace{0.5\baselineskip}
\begin{theorem} \label{thm:ExpConv}
    Consider a group of circular shaped robots modeled with single-integrator dynamics \eqref{eq:AgentDynamics} and having a graph topology $ \mathcal{G} $ such that the desired formation is minimally and infinitesimally rigid in the distance rigidity sense.
    Let $ e\BR{0} $ be such that it is in the compact and invariant set $ \Omega_{\beta} $ \eqref{eq:OmegaBeta}. Then $ e = \mathbb{0}_{m} $ is a locally exponential stable equilibrium point of the error dynamics \eqref{eq:ErrorDynamics}.
\end{theorem}

\vspace{0.5\baselineskip}
\begin{proof}
    The proof can be divided into three main stages. 
    \textit{First}, we consider the asymptotic stability of the origin $ e = \mathbb{0}_{m} $. 
    The set $ \Omega_{\beta} $ has the property of being compact and positive invariant.
    In addition, the value for $ \beta $ can be chosen such that for every vector $ e \in \Omega_{\beta} $, the formation is minimally and infinitesimally rigid in the distance rigidity sense, and close to the target formation.
    Due to our choice of $ \beta $, we have that $ R_{\text{dist}} $ has full row rank.
    {Since} $ R_{\text{angle}} = D\BR{d} R_{\text{dist}} $ {and} $ D\BR{d} $ positive definite, also $ R_{\text{angle}} $ has full row rank. 
    This in turn implies $ F\BR{e} = R_{\text{angle}} R_{\text{angle}}^{\top} $ is positive definite. 
    Let $ \lambda $ be the minimal eigenvalue of the matrix $ F\BR{e} $ in $ \Omega_{\beta} $, i.e., 
    $ \lambda = \min_{e \in \Omega_{\beta}} \textbf{eig} \BR{F\BR{e}} > 0 $.
    It follows from \eqref{eq:OverallPotFuncDer} that  
    \begin{equation} \label{eq:OverallPotFuncDerV}
        \dot{V}\BR{e} 
            = -v^{\top}\BR{e} F\BR{e} v\BR{e}
            \leq - \lambda \norm{v\BR{e}}^{2}
    \end{equation}
    holds. 
    The value $ \dot{V}\BR{e} $ is negative definite for all $ e \in \Omega_{\beta} \setminus \CBR{\mathbb{0}_{m}} ${; therefore,} local asymptotic stability of the origin is attained. 
    
    \vspace{0.5\baselineskip}
    \textit{Next}, we aim to show the following two inequalities as is done in \cite{Sun2016}:
    \begin{equation} \label{eq:Inequalities}
        \begin{aligned}
            c_{1} \norm{e}^{2} 
            \leq V\BR{e} \leq c_{2} \norm{e}^{2}
            ; \quad
            \norm{v\BR{e}}^{2} 
            \geq \rho \norm{e}^{2}
            ,
        \end{aligned}
    \end{equation}
    with $ c_{1} $, $ c_{2} $, and $ \rho $ be{ing} positive constants that we need to determine. 
    These inequalities facilitate the proof to exponential stability of the origin. 
    To this end, recall the overall potential function $ V\BR{e} $ 
    \begin{equation} \label{eq:VariableGradientPotFunc}
        V\BR{e} 
            = \sum_{k \in \mathcal{K}} V_{k}\BR{e_{k}} 
            = \sum_{k \in \mathcal{K}} \int_{0}^{e_{k}} v_{k}\BR{s} \text{ d}s.
    \end{equation}
    Within the set $ \Omega_{\beta} $, we can find a value for $ \delta $ such that 
    \begin{equation} \label{eq:HypercubeDelta}
        \mathcal{H}_{\delta} = \CBR{e \in \Omega_{\beta} \, \rvert \, \ABS{e_{k}} \leq \delta, \, k \in \mathcal{K}}.
    \end{equation}
    By Lemma 3.2 in \cite{Khalil2001}, we have the function $ v_{k}\BR{e_{k}} $ is Lipschitz continuous in $ \mathcal{H}_{\delta} $.
    In addition, the function $ k_{k}\BR{e_{k}} $ is positive within the set $ \Omega_{\beta} ${,} and thus {also}  in the subset $ \mathcal{H}_{\delta} $. 
    The remainder of the proof for obtaining the positive constants $ c_{1} $, $ c_{2} $, and $ \rho $ of \eqref{eq:Inequalities} follows closely \cite{Sun2016} and {for this reason, it is}  omitted.
    
    \textit{Finally}, we can show exponential stability of the origin as a result of the previous two steps. 
    Substituting \eqref{eq:Inequalities} in \eqref{eq:OverallPotFuncDerV}, we obtain
    \begin{equation} \label{eq:OverallPotFuncDerV3}
        \dot{V}\BR{e} \leq - \lambda \norm{v\BR{e}}^{2} \leq - \lambda \rho \norm{e}^{2}.
    \end{equation}
    By Theorem 4.10 in \cite{Khalil2001}, we can conclude that the origin is exponential{ly} stable in $ \mathcal{H}_{\delta} $. 
    The error norm can be shown to be bounded by an exponential decreasing function as
    \begin{equation} \label{eq:BoundedError}
        \begin{aligned}
            \norm{e\BR{t}} 
                \leq {\BR{\frac{c_{2}}{c_{1}}}}^{\frac{1}{2}} \norm{e\BR{0}} \operatorname{exp}\BR{- \frac{\gamma}{2} t},
        \end{aligned}
    \end{equation}
    with $ \gamma = \frac{\lambda \rho}{c_{2}} $.
    This concludes the proof.
\end{proof}

\begin{figure*}
    \centering
    {
    \subfigure[Robot trajectories]
    {
    \includegraphics[width=0.230\textwidth]{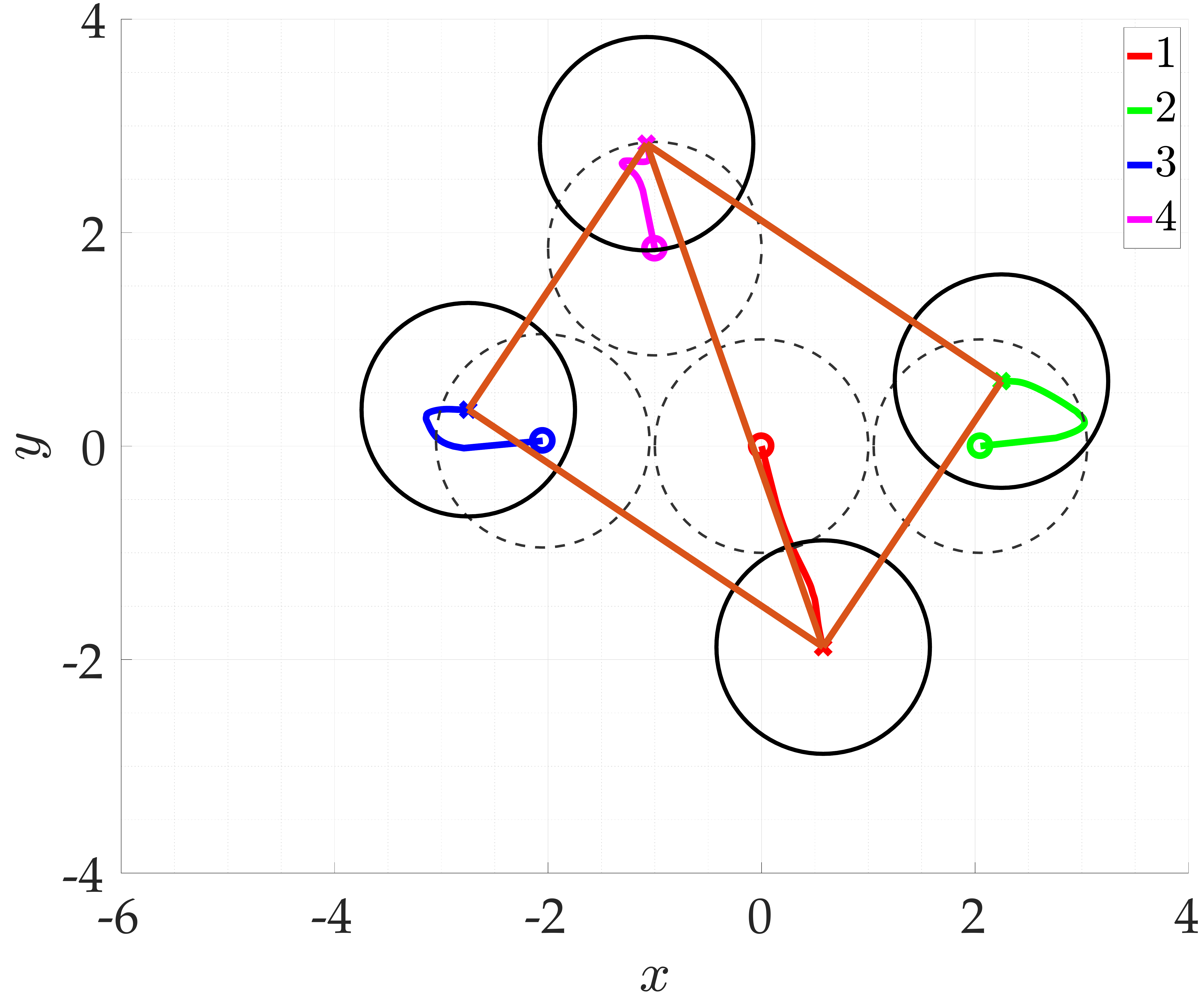} % 0.2125
    \label{fig:Sim-Agent-Evol}
    }
    \hfill
    \subfigure[Inter-center distances]
    {\includegraphics[width=0.35\textwidth]{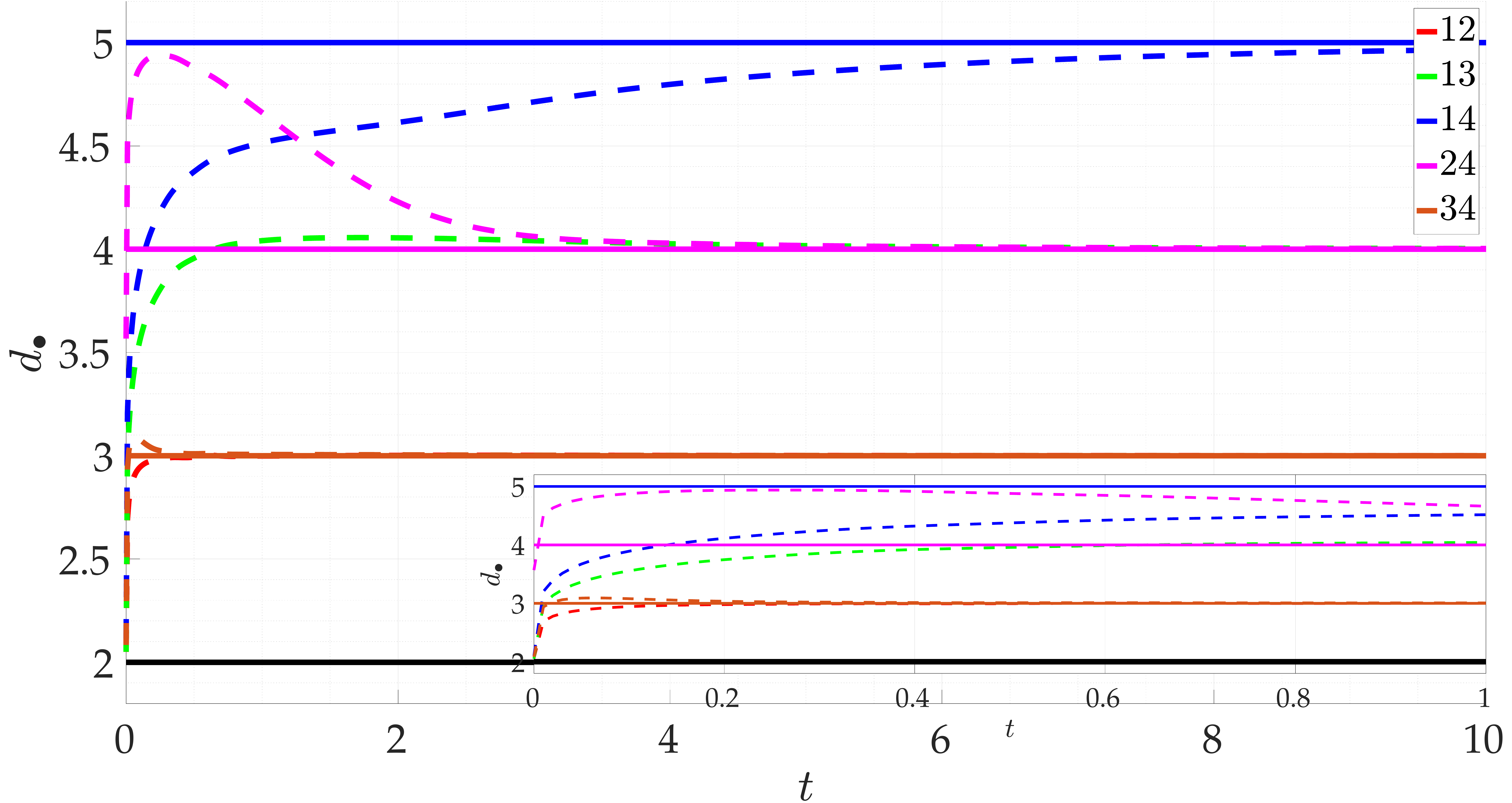} % 0.365
    \label{fig:Sim-Distance-Evol}
    }
    \hfill
    \subfigure[Internal angle errors]
    {\includegraphics[width=0.35\textwidth]{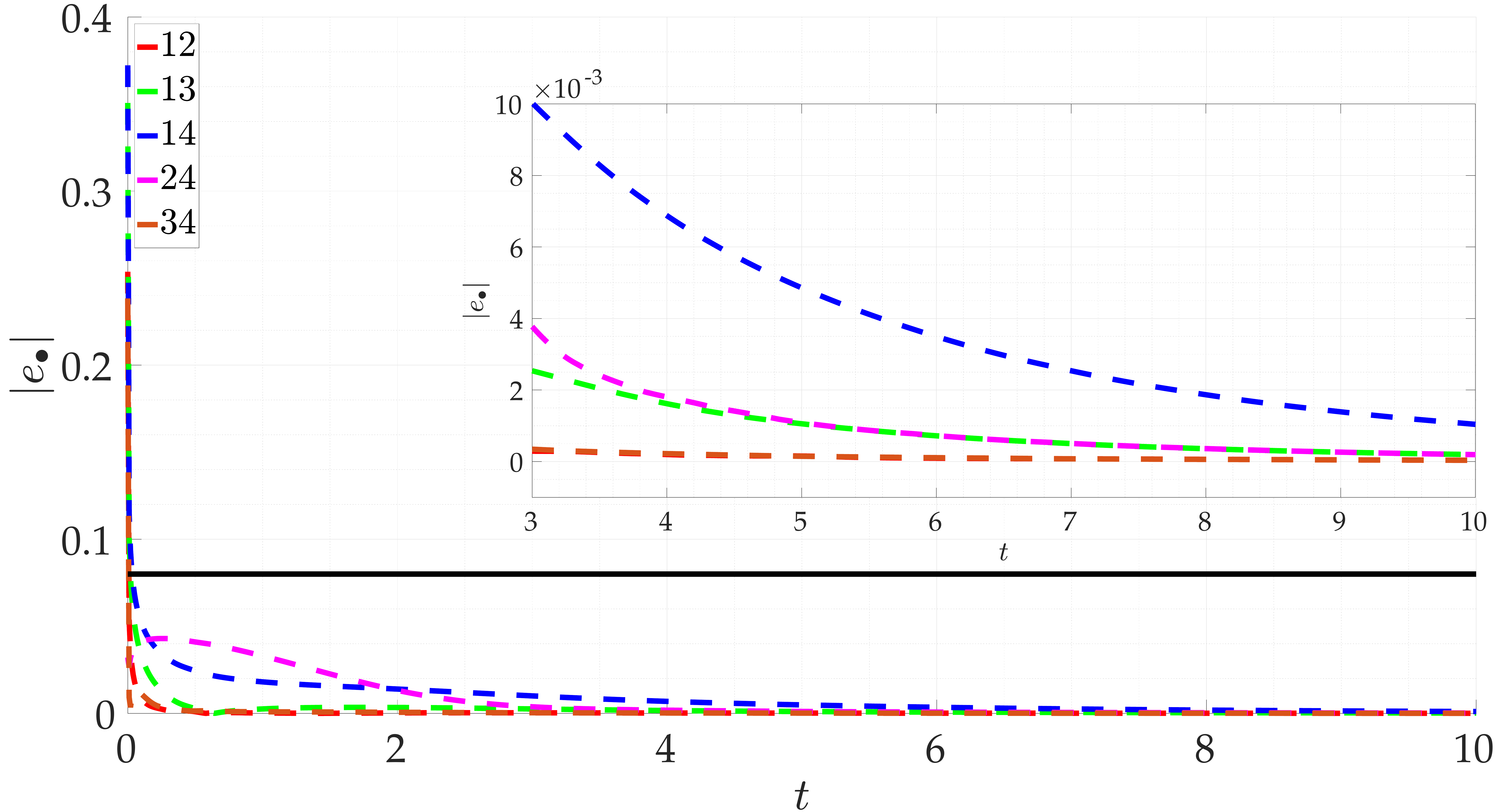}
    \label{fig:Sim-AngleError-Evol}
    }
    }
\caption{Simulation with a team of $ 4 $ circular mobile robots having radii $ r = 1 $. On the left panel, we have the robot trajectories; dashed circles represent initial configuration while solid circles are final robot positions. The solid lines are the robot center trajectories. In the center panel, the convergence of the distances $ d_{ij} $ (dashed) to their desired values $ d_{ij}^{\star} $ (solid) is depicted. The black solid line represents $ d_{\min} = 2 $ between the robots. The right panel shows the convergence of the internal angle errors. The black solid line depicts the value $ b = 0.08 $ for the hypercube $ \mathcal{H}_{b} $.}
    \label{fig:Sim-Result}
\end{figure*}

\subsection{Equilibrium Sets}
\label{sec: shapes}
Theorem \ref{thm:ExpConv} concerns the local exponential convergence of the formation control system to the desired formation shape. 
In general, the set of equilibrium points of the mobile robots can be given by 
$ \mathcal{W} \coloneqq \CBR{p \in \RN^{2n} \, \rvert \,  R_{\text{angle}}^{\top}v\BR{e} = \mathbb{0}_{2n}} $.
The set of \textit{correct} formation shapes can be given by 
$ \mathcal{W}_{\text{c}} \coloneqq \CBR{p \in \RN^{2n} \, \rvert \,  e = \mathbb{0}_{m}} $ 
while the set of \textit{incorrect} formation shapes is $ \mathcal{W}_{i} \coloneqq \mathcal{W} \setminus \mathcal{W}_{c} $.
Considering the target formation shape is minimally and infinitesimally rigid{,} we can conclude that the formation shapes in $ \mathcal{W}_{i} $ are not infinitesimally rigid{, since}  the null space of $ R_{\text{angle}}^{\top} $ {also} consists of a non-trivial vector $ v\BR{e} \neq \mathbb{0}_{m} $. {As in distance-based control, the set $\mathcal{W}_{i}$ includes configurations where all the robots' center are collinear.} {Moreover,}  we can obtain the following on the equilibrium set of the $ p $-dynamics and the $ e $-dynamics:

\vspace{0.5\baselineskip}
\begin{lemma} \label{lem:EquilibriumSets}
    The equilibrium sets of the error system \eqref{eq:ErrorDynamics} is the same as the equilibrium sets of the closed-loop formation {control} system \eqref{eq:OverallClosedLoopExp}. 
\end{lemma}

%%%%

\vspace{0.5\baselineskip}
\begin{proof}
    {
    Since $ \dot{e}\BR{t} = R_{\text{angle}} \dot{p}\BR{t} $, obviously $ \dot{p}\BR{t} = \mathbb{0}_{2n} 
    \implies \dot{e}\BR{t} = \mathbb{0}_{m} $. 
    It remains to show $ \dot{e}\BR{t} = \mathbb{0}_{m} \implies \dot{p}\BR{t} = \mathbb{0}_{2n} $. 
    Assume $ \exists \dot{p}\BR{t} \neq \mathbb{0}_{2n} $ such that $ \dot{p}\BR{t} \in \NULL{R_{\text{angle}}} $ holds.
    From \eqref{eq:OverallClosedLoopExp}, we also have $ \dot{p}\BR{t} \in \COL{R_{\text{angle}}^{\top}} $. 
    Since, $ \NULL{R_{\text{angle}}} \perp \COL{R_{\text{angle}}^{\top}} $, we obtain $ \NULL{R_{\text{angle}}} \, \cap \, \COL{R_{\text{angle}}^{\top}} = \CBR{\mathbb{0}_{2n}} $, contradicting the assumption $ \dot{p}\BR{t} \neq \mathbb{0}_{2n} $.
    This concludes the proof.
    }
\end{proof}

\section{NUMERICAL EXAMPLE} \label{sec:NumericalExample}
\subsection{Simulation Setup}
{
We apply the proposed control law to a team of $ 4 $ circular robots with radii $ r = 1 $. 
The collective goal is to form a rectangular shape with the inter-center distances given as $ d_{12}^{\star} = d_{34}^{\star} = 3 $, $ d_{13}^{\star} = d_{24}^{\star} = 4 $, and $ d_{14}^{\star} = 5 $.
Using \eqref{eq:CosineRobot_ij}, we obtain 
$ \cos \theta_{12}^{\star} = \cos \theta_{34}^{\star} = 0.7778 $, $ \cos \theta_{13}^{\star} = \cos \theta_{24}^{\star} = 0.8750 $, and $ \cos \theta_{14}^{\star} = 0.9200 $.
The initial configuration, depicted as dashed circles in Fig. \ref{fig:Sim-Agent-Evol}, has center positions 
$ p_{1}\BR{0} = \left[0, \, 0 \right]^{\top} $, $ p_{2}\BR{0} = \left[2.05, \, 0 \right]^{\top} $, $ p_{3}\BR{0} = \left[-2.05, \, 0.05 \right]^{\top} $, and $ p_{4}\BR{0} = \left[-1, \, 1.85 \right]^{\top} $.
Using this initial configuration, we can 
illustrate the collision avoidance feature of the proposed control law and the convergence to the desired formation shape, even though $ p\BR{0} \not\in \mathcal{H}_{b} $.
We can obtain $ b = 0.08 $, and set the gain $ K = 50 $ for speeding up the convergence. 
}

\subsection{Simulation Results}
{
The trajectories of the robots are depicted in Fig. \ref{fig:Sim-Agent-Evol}. In addition, the inter-center distances and the internal angle errors between the robots are given in Figs. \ref{fig:Sim-Distance-Evol} and \ref{fig:Sim-AngleError-Evol}, respectively. 
Let us focus on robot $ 2 $, the green robot in Fig. \ref{fig:Sim-Agent-Evol}.
It has the neighboring robots $ 1 $ (red robot) and $ 4 $ (magenta robot).
From the figure, we observe that since robots $ 2 $ and $ 1 $ are close to each other initially, robot $ 2 $ quickly moves away from robot $ 1 $, and almost attains the desired constraint with robot $ 1 $.
However, due to this motion, its distance to the neighboring robot $ 4 $ has increased to about $ 4.9 $. 
This can also be observed from Fig. \ref{fig:Sim-AngleError-Evol}, where we see an increase in the magenta colored signal representing error $ \ABS{e_{24}} $.
Since robot $ 2 $ is now sufficiently far from robot $ 1 $, it then tries to satisfy the internal angle constraint with robot $ 4 $ as can be observed in both Figs. \ref{fig:Sim-Distance-Evol} and \ref{fig:Sim-AngleError-Evol} . 
By zooming in on Fig. \ref{fig:Sim-AngleError-Evol}, we can observe exponential convergence of the error signals starting around $ t = 3 $s. 
All the error signals are then well below the threshold value of $ b = 0.08 $.
}

%%%

\section{CONCLUSIONS} \label{sec:Conclusions}
In this letter, we {have} solved the formation control problem for circular mobile robots {subjected to} internal angle constraints.
A gradient-descent control law {requiring} only  relative bearing measurements for implementation {has been}  proposed. 
This control law {enjoys} local exponential convergence for the error dynamics and {ensures} collision avoidance between {neighboring} robots.

%%%%%%%%%%%%%%%%%%%%%%%%%%%%%%%%%%%%%%%%%%%%%%%%%%%%%%%%%%%%%%%%%%%%%%%%%%%%%%%%

\bibliographystyle{IEEEtran}
\bibliography{V2-LCSS-CDC2020}
% \printbibliography

%%%%%%%%%%%%%%%%%%%%%%%%%%%%%%%%%%%%%%%%%%%%%%%%%%%%%%%%%%%%%%%%%%%%%%%%%%%%%%%%
% % EXTRA

\end{document}